\documentclass[aps,prb,twocolumn,amsmath,amssymb,superscriptaddress,floatfix]{revtex4}
\usepackage{graphicx}
\usepackage{bm}
\usepackage[usenames]{color}
\bibstyle{apsrev.bib}

\newcommand{\be}{\begin{equation}}
\newcommand{\ee}{\end{equation}}
\newcommand{\beqn}{\begin{eqnarray}}
\newcommand{\eeqn}{\end{eqnarray}}

\begin{document}

\title{Renormalization group study of the two-dimensional random transverse-field Ising model}
\author{Istv\'an A. Kov\'acs}
\email{ikovacs@szfki.hu}
\affiliation{Department of Physics, Lor\'and E\"otv\"os University, H-1117 Budapest,
P\'azm\'any P. s. 1/A, Hungary}
\affiliation{Research Institute for Solid State Physics and Optics,
H-1525 Budapest, P.O.Box 49, Hungary}
\author{Ferenc Igl\'oi}
\email{igloi@szfki.hu}
\affiliation{Research Institute for Solid State Physics and Optics,
H-1525 Budapest, P.O.Box 49, Hungary}
 \affiliation{Institute of Theoretical Physics,
Szeged University, H-6720 Szeged, Hungary}
\date{\today}

\begin{abstract}
The infinite disorder fixed point of the random transverse-field Ising model is expected to control the critical behavior of a large class of random quantum and
stochastic systems having an order parameter with discrete symmetry. Here we study the model on the square lattice with a very efficient numerical implementation
of the strong disorder renormalization group method, which makes us possible to treat finite samples of linear size up to $L=2048$. We have calculated
sample dependent pseudo-critical points and studied their distribution, which is found to be characterized by the same shift and width exponent: $\nu=1.24(2)$.
For different types of disorder the infinite disorder fixed point is shown to be characterized by the same set of critical
exponents, for which we have obtained improved estimates: $x=0.982(15)$ and $\psi=0.48(2)$. We have also studied the scaling behavior of the magnetization in
the vicinity of the critical point as well as dynamical scaling in the ordered and disordered Griffiths phases.
\end{abstract}


\maketitle

\section{Introduction}
\label{sec:intro}
The random transverse-field Ising (RTFI) model is the prototype of random quantum systems\cite{qsg} having a quantum critical point at zero temperature\cite{sachdev}.
This model has experimental realizations\cite{experiment} and there is a large amount of theoretical work, which aims to clarify the properties of the random critical point.
It is expected that basic features of the critical behavior are demonstrated in the one-dimensional ({\it 1d}) model and therefore 
most of the theoretical studies are performed in {\it 1d}. After early works by McCoy and others\cite{mccoywu,
shankar} Fisher\cite{fisher} has used a renormalization group (RG)
framework to obtain several presumably exact results about its critical properties. It has been shown that the critical properties of the {\it 1d} model are governed by a
so called infinite disorder fixed point (IDFP), in which the strength of disorder grows without limit during renormalization\cite{danielreview}. As a consequence disorder
fluctuations are dominated over quantum fluctuations and the approximations used in the RG approach become exact at the critical point. The IDFP of the RTFI model
is shown to govern the critical properties of another random quantum systems having an order parameter with discrete symmetry\cite{senthil,carlon-clock-at}.
Furthermore this fixed point is found
to be isomorphic with that of several {\it 1d} stochastic models, such as the Sinai walk\cite{rgsinailong}, the random contact process\cite{hiv} or the
random exclusion process\cite{igloipartial}.

Comparatively less results are available about the critical behavior of the RTFI model in higher dimensions, which are almost exclusively restricted to {\it 2d}.
By now different numerical studies are in favor of the conclusion, that also in {\it 2d} the critical behavior is controlled by an IDFP. In this respect we
mention different numerical implementations\cite{motrunich00,lin00,karevski01,lin07,yu07,ladder,pich98} of the strong disorder renormalization
group (SDRG) method, as well quantum Monte Carlo simulations\cite{pich98}. These results are in agreement with recent simulation studies of the {\it 2d} random contact process\cite{vojta09}, which is
expected to belong to the same universality class. Also the {\it 2d} random walk in a self-affine random potential\cite{selfaff}
could be
related to the {\it 2d} RTFI model. As far as the numerical estimates of the critical exponents in {\it 2d} are concerned, these
contain quite large errors, for a summary of the estimates see Ref.[\onlinecite{ladder}]. In simulation studies these are connected to the logarithmically slow critical relaxation, whereas in the SDRG method the errors
has mainly finite-size origin the typical linear size of the largest systems being about $L=128-160$. Also the type of
disorder used in the calculations has an influence on the error of the results. Due to these numerical limitations there are no quantitative results in {\it 3d}, although
it is very probable that the random critical point is an IDFP in this case, too\cite{motrunich00}.

In this paper we are going to revisit the problem of the critical behavior of the {\it 2d} RTFI model. Like other studies we use
numerical implementation of the SDRG method, however we have developed a very efficient algorithm, which make us possible
to treat systems as large as $L=2048$. In this way the number of sites in our samples are several hundred larger,
than in previous studies. Comparing with earlier SDRG investigations our study has several different features. i) We define
and calculate finite-size pseudo-critical points and study their distribution. ii) We obtain accurate estimates for the
true critical point of the model, calculate effective, size-dependent critical exponents and study their extrapolation.
iii) We consider different forms of the disorder and study the universality of the critical exponents as well as
the scaling functions. iv) We also study
scaling outside the critical point, as well as dynamical scaling in the disordered and ordered Griffiths phases.

The structure of the rest of the paper is the following. The model and the SDRG method is presented in Sec. \ref{sec:model}.
The basic features of
the computer algorithm are discussed in Sec. \ref{sec:impl}.
In Sec. \ref{sec:TC} we describe how finite-size transition points are defined and calculated within the
frame of the SDRG method. We study their distribution and analyze the shift of the mean value as well as
the width as a function of the
size of the system. In Sec. \ref{sec:crit} critical exponents are extracted through finite-size scaling and
the scaling behavior outside the critical point is analyzed. We also study dynamical scaling in the disordered and in
the ordered
Griffiths phases. Our results are discussed in the final section, whereas some details of the computer algorithm are
presented in the Appendix.

\section{Model and the SDRG method}
\label{sec:model}

\subsection{Random transverse-field Ising model}

We consider the RTFI model defined by the Hamiltonian:
\be
{\cal H} =
-\sum_{\langle ij \rangle} J_{ij}\sigma_i^x \sigma_{j}^x-\sum_{i} h_i \sigma_i^z
\label{eq:H}
\ee
in terms of the Pauli-matrices, $\sigma_i^{x,z}$. Here $i$ and $j$ are sites of a square lattice and the first sum runs over nearest neighbors. The $J_{ij}$ couplings and the $h_i$ transverse fields are independent random numbers, which are taken from the
distributions, $p(J)$ and $q(h)$, respectively.

Here we use two different type of distributions, which have both the same uniform distribution of the couplings:
\be
p(J)=\Theta(J)\Theta(1-J)\;, 
\ee
$\Theta(x)$ being the Heaviside step-function. For the 'box-$h$' disorder we have:
\be
q(h)=\frac{1}{h_b}\Theta(h)\Theta(h_b-h)\;, 
\label{box_h}
\ee
whereas for the 'fixed-$h$' model we have a constant transverse field:
\be
q(h)=\delta(h_f-h)\;.
\label{fixed_h}
\ee

In the following we use the logarithmic transverse-field $\theta=\ln{h_b}$ or $\theta=\ln{h_f}$ to characterize the
system. In the thermodynamic limit, $L \to \infty$, the system in Eq.(\ref{eq:H}) displays a paramagnetic phase, for $\theta>\theta_c$, and a ferromagnetic phase, for $\theta<\theta_c$. In between there is a random quantum critical point at $\theta=\theta_c$. The quantum control-parameter is defined as $\delta=\theta-\theta_c$.

\subsection{Strong disorder renormalization group method}

\begin{figure}[h!]
\begin{center}
\includegraphics[width=2.8in,angle=0]{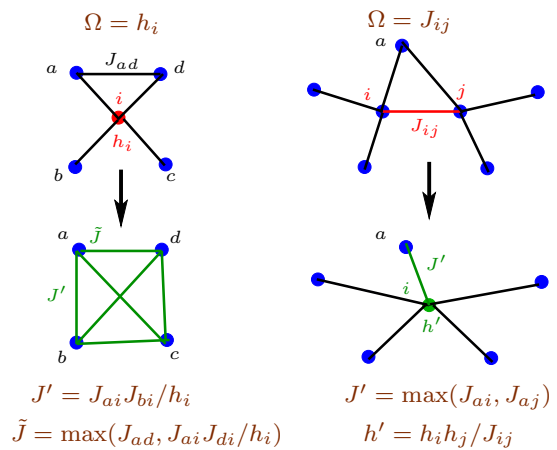}
\end{center}
\caption{
\label{fig_1} 
Illustration of the decimation steps of the strong disorder renormalization group method in higher dimensions.
}
\end{figure}

Here we use the SDRG method\cite{im}, which has been introduced by Ma,
Dasgupta and Hu\cite{mdh} and later developed by D. Fisher\cite{fisher} and others. During the method, the largest local term in the Hamiltonian (which defines the energy-scale, $\Omega$, at the given RG step) is successively eliminated and at the same time new terms are generated between remaining sites by second-order perturbation method. The procedure is sketched in Fig. \ref{fig_1} for a higher dimensional system. If the largest term is a coupling (see the right panel of Fig. \ref{fig_1}), say $\Omega=J_{i,j}$ connecting sites $i$ and $j$, then the two sites involved form a spin cluster with an effective moment $\mu'=\mu_i+\mu_j$ (initially $\mu_i=1~\forall i$) in
the presence of an effective transverse field: $h' \approx h_i h_j/J_{i,j}$. The renormalized value
of the coupling of the cluster to a site $a$
is given by the 'maximum rule' $\max{[J_{a,i},J_{a,j}]}$.
On the other hand, if the largest local term is a transverse-field (see the left panel of Fig. \ref{fig_1}), say $\Omega=h_i$, then site $i$ is eliminated and effective couplings are generated
between each pairs of spins, among the neighbors of $i$. If $a$ and $b$ are neighboring spins to $i$,
then the generated coupling is given by: $J_{a,b}' \approx J_{a,i} J_{b,i}/h_{i}$. If the sites $a$ and $b$ are
already connected by a coupling, $J_{a,b} \ne 0$, than the renormalized coupling is
given by the 'maximum rule' as $\max{[J_{a,b},J_{a,b}']}$. The use of the maximum rule is justified if the renormalized couplings have a very broad distribution, which is indeed the case at the IDFP. We shall see that with the maximum rule
the numerical algorithm can be simplified.
At each step of the renormalization one site is eliminated and the energy scale is continuously lowered. For a finite system the renormalization is stopped at the last site, where we keep the energy-scale and the total moment as well as the structure of the clusters.

\subsection{Scaling at the infinite disorder fixed point}

At the IDFP the distribution of the effective couplings and that of the transverse fields becomes broader and broader during the renormalization\cite{danielreview,im}. Considering the ratio of two effective 
terms in the Hamiltonian at a given stage of the SDRG it will tend either to infinity or to zero. This indicates that the disorder is infinitely strong and the perturbative
results during the RG are exact. Qualitatively the log-energy scale, $\ln \Omega$, scales with the linear size of the system size, $L$:
\be
\ln{ \left(\Omega_0/\Omega\right)} \sim L^{\psi}\;,
\label{psi}
\ee
where $\Omega_0$ denotes a reference energy scale.
The average spin-spin correlation function is defined as $G(r)=\overline{\langle \sigma_i^x \sigma_{i+r}^x}\rangle$, where $\langle \dots \rangle$ denotes the ground-state average and $\overline{(\dots)}$ stands for the averaging over quenched disorder.
The asymptotic value of the correlation function defines the magnetization, $m$, in the system:
\be
\lim_{r \to \infty} G(r)=m^2\;,
\label{m_def}
\ee
and $m>0$ in the ferromagnetic phase and $m=0$ in the paramagnetic phase.
The connected correlation function, $\tilde{G}(r)=G(r)-m^2$, in the vicinity of the critical point behaves as:
\be
\tilde{G}(r)\sim r^{-2x}\exp(-r/\xi)\;,
\label{Gr}
\ee
where the correlation length, $\xi$, is divergent at the critical point as:
\be
\xi \sim \left|\delta\right|^{-\nu}\;.
\label{xi}
\ee
Thus at $\delta=0$ there is a power-law decay of the correlations,
which is related to the fractal structure of the spin clusters. Indeed,
the average cluster moment, $\mu$, is related to the energy-scale, $\Omega$ as:
\be
\mu \sim \left[\ln (\Omega_0/\Omega)\right]^{\phi}\;,
\label{phi}
\ee
and can be expressed also with the size:
\be
\mu \sim L^{d_f}\;.
\label{mu}
\ee
Here the fractal dimension of the cluster, $d_f$, is related to the other exponents as:
\be
d_f=\phi \psi=d-x\;.
\label{d_f}
\ee
In {\it 1d} the critical exponents are exactly known\cite{mccoywu,fisher}:
\be
\psi=\frac{1}{2},\;\quad \phi=\frac{\sqrt{5}+1}{2},\; \quad \nu=2,\; \quad x=\frac{3-\sqrt{5}}{4}.
\ee

\section{Implementation of the SDRG method}
\label{sec:impl}
\subsection{Problems with the na\"{\i}ve implementation}
The SDRG decimation rules are very simple and it is straightforward to implement the method numerically.
In higher dimensions, however, the topology of the lattice is changing during renormalization, which could
result in considerable increase of the computational time. More dangerous steps in
this respect are the $h$-decimations, during which numerous new bonds are generated and as a result sites
with large number of links are formed. In this way, after a na\"{\i}ve implementation of the method,
the system is transformed into an almost complete graph
and the subsequent $h$-decimations, generating new links between practically all pair of sites, will be very slow.
For a system with $N$ sites this algorithm would need $\mathcal{O}(N^3)$ time.

Using the maximum rule in the approach, however, offers two ways to speed up the procedure\cite{note}. First, one can notice
that the renormalization trajectory is not unique in this case. There are terms in the Hamiltonian, which are called as
``local maxima'' and which can be decimated independently. Thus one should not follow the ``decimation of the
largest term in each step'' principle, instead we are going to optimize the time of the renormalization trajectory,
which goes over in some order of the local maxima. The second consequence of the maximum rule is, that
a large number of bonds will never be participating in the renormalization process. These latent
bonds can be deleted from the list of edges without
consequences. The latent bonds are in such a
local environment, that after decimating a nearby site or bond a stronger new coupling is
generated to the same edge, thus the original bond disappears without participating in the renormalization.
Filtering out these irrelevant bonds will result in a considerable improvement of the algorithm.
In the following we
discuss the properties of the local maxima and the optimal RG trajectory as well as the main features of the filtering process.

\subsection{Local maxima and the optimal RG trajectory}

A local maximum in the set of couplings and transverse fields is such a term, which is larger (not smaller) than any of its
neighboring terms. Considering a coupling $J_{ij}$ is a local maximum, provided $J_{ij} \ge h_i$, $J_{ij} \ge h_j$, and
$J_{ij}\ge J_{ik},~\forall k$, as well as $J_{ij}\ge J_{lj},~\forall l$. Similarly a transverse field, $h_i$,
is a local maximum, if $h_i \ge J_{ij},~\forall j$. It can be shown that the local maxima can be decimated independently,
the renormalization performed in any sequence gives the same final result.

\subsection{Filtering out irrelevant bonds}
The principle, that some bonds are irrelevant and does not modify the renormalization procedure has been first
realized by Kawashima\cite{kawashima}. He has also suggested a criterion to identify the
irrelevant bonds which are then deleted from the graph. This filtering procedure as illustrated in Fig. \ref{fig_2}
is used in a few {\it 2d}
numerical works\cite{karevski01,lin07,ladder}. In Appendix \ref{kawa_app} we give the proper definition of the filtering
criterion and prove it.

\begin{figure}[h!]
\begin{center}
\includegraphics[width=2.8in,angle=0]{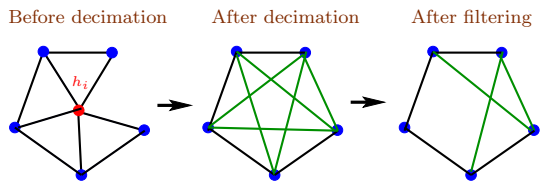}
\end{center}
\caption{
\label{fig_2} 
Illustration of the filtering criterion.
}
\end{figure}

The use of the filtering criterion for a site with $k$ neighbors needs typically $\mathcal{O}(k^3)$
time, since $\mathcal{O}(k^2)$ triangles have to be checked. However, the application of the filtering
makes not too much reduction of the computational time, if the bond is in a general position, where not too many new terms are generated
after consecutive RG steps. There are, however, bonds in ``dangerous positions'', having at least one of the neighboring
transverse fields as a local maximum. Decimating this local maximum many
new terms are generated and therefore the use of prefiltering, which checks the bonds of the decimated spin just before its decimation is very effective, which typically needs only $\mathcal{O}(k^2)$ time.

\subsection{Basic steps in the numerical SDRG method}

In order to obtain an efficient implementation of the SDRG method we combine
the optimal selection of the RG trajectory with the filtering algorithm. In our method we use the following
steps.
\begin{enumerate}
 \item We check all terms of the Hamiltonian and select the local maxima. We make two lists, one for
the couplings and one for the transverse fields.
\item We decimate every coupling, which is on the list of local maxima.
In this step the order of the decimation is arbitrary. After decimating the original couplings
new terms are generated, among which there are new local maxima. We include those into the list
of local maxima and at the same time we decimate the new couplings from this list. We
repeat this step, until the list of local maximum couplings is empty.

\item If all or all but one the transverse fields are local maxima we select the smallest one,
delete all the rest sites and END the iteration. Otherwise
we select a transverse field from the list of local maxima having a small (or the smallest)
degree, which is defined as the number of edges
to the given site and filter for
the neighboring bonds. We decimate this transverse field and check the generated new terms for local maxima.
If the list of local maximum couplings is not empty, we go to step 2. Otherwise we go to step 3.
\end{enumerate}
Using the selection rule in the second part of step 3. will ensure that the average degree of the sites will
not be large. In this way we prevent the formation of too connected clusters, the decimation
of which being time consuming. In step 3. making the filtering before decimation (prefiltering) will
ensure that the dangerous bonds are deleted.

This algorithm in {\it 2d} works typically in $\mathcal{O}(L^2 \ln L)$ time in a $L \times L$ system near
the critical point.
With our method an $L=128$ sample is renormalized in $\sim 1$ second, whereas for
$L=1024$ the typical time is $\sim 1.5$ minutes (in a 2.4GHz processor).

\section{Finite-size critical points}
\label{sec:TC}
\subsection{Scaling of pseudo-critical points}
In the study of the critical behavior of random systems it is very important to find an accurate
estimate of the location of the critical point. The quality of the estimate of $\theta_c$ will influence
the error of the calculated critical exponents and scaling functions. In a random sample of linear size, $L$,
one can generally define finite-size pseudo-critical points\cite{psz,domany,aharony,paz2,mai04,PS2005},
$\theta_c(L)$, which are usually given as the position of
the maximum of some physical quantity, which is divergent in the thermodynamic limit (c.f. susceptibility) at $\theta_c$.
The distribution of $\theta_c(L)$ provides important information about the scaling behavior at the fixed point
of the system\cite{aharony}. In particular one studies the shift of the average value, $\overline{\theta_c}(L)$, which is expected
to scale as:
\be
\left|\theta_c-\overline{\theta_c}(L)\right| \sim L^{-1/\nu_s}\;,
\label{nu_s}
\ee
with the shift exponent, $\nu_s$. Similarly, one measures the width of the distribution, $\Delta \theta_c(L)$, which
behaves for large-$L$ as:
\be
 \Delta \theta_c(L) \sim L^{-1/\nu_w}\;,
\label{nu_w}
\ee
where $\nu_w$ denotes the width exponent. According to renormalization group theory \cite{aharony} for a classical random system
with {\it relevant} disorder\cite{harris} the critical behavior is controlled by a conventional random fixed point, with
$\nu_s=\nu_w=\nu$, where $\nu \ge 2/d$\cite{ccfs} is the correlation length critical exponent of the system.

\subsection{Identification of pseudo-critical points}
\label{sec:pseudo-cr-pt}
In a random quantum system one has to use another methods to locate pseudo-critical points\cite{ilrm}. One method, which is
well suited to the SDRG approach is the doubling method\cite{PS2005}, which has been used for chains\cite{ilrm}
and for ladders\cite{ladder} of the
RTFI model. In the doubling procedure in {\it 1d} (or quasi-{\it 1d}) geometry one considers a random sample ($\alpha$) of
length, $L$, and makes a duplicated sample ($2\alpha$) of length $2L$ by joining two copies of ($\alpha$). Using
the SDRG method one calculates some physical quantity (magnetization or gap)
in the original and in the replicated sample, which is denoted by $f(\alpha,L)$ and $f(2\alpha,2L)$, respectively,
and study their ratio, $r(\alpha,L)=f(2\alpha,2L)/f(\alpha,L)$, as a function of the control parameter, $\theta$.
At $\theta=\theta_c(\alpha,L)$ this ratio has a sudden jump, which is identified with the pseudo-critical point of
the sample. We note that the actual value of $\theta_c(\alpha,L)$ is practically independent of the physical
quantity we considered\cite{ladder}, since this singularity is connected to the topology of clusters produced by the SDRG approach.

\begin{figure}[h!]
\begin{center}
\includegraphics[width=1.8in,angle=0]{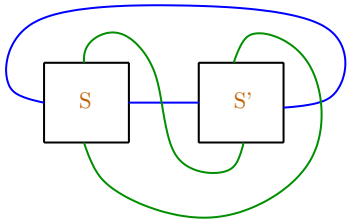}
\end{center}
\caption{
\label{fig_3} 
Illustration of the boundary conditions used in the doubling procedure in higher dimensions.
}
\end{figure}

Having this observation in mind we can generalize the doubling method for two (and higher) dimensions. 
In {\it 2d} we glue together two identical square-shaped samples at the boundaries as indicated in Fig. \ref{fig_3}.
Then we renormalize the duplicated sample and calculate the structure of the connected clusters, among which there
might be such, which have sites (in equivalent positions) in both replicas. These sites are correlated and the
fraction of these correlated sites defines the correlation function between the replicas. By increasing the control
parameter, $\theta$, the replica correlation function is decreasing and at a well defined value, $\theta_c(L)$, it
suddenly jumps to zero. We consider $\theta_c(L)$ as the pseudo-critical point of the given sample. It is easy to see
that this definition is equivalent to the previously used criterion in {\it 1d}, furthermore it is straightforward to
generalize it to three- and higher dimensions.

\subsection{Numerical results}
\subsubsection{Distribution of pseudo-critical points}
\begin{figure}[h!]
\begin{center}
\includegraphics[width=3.4in,angle=0]{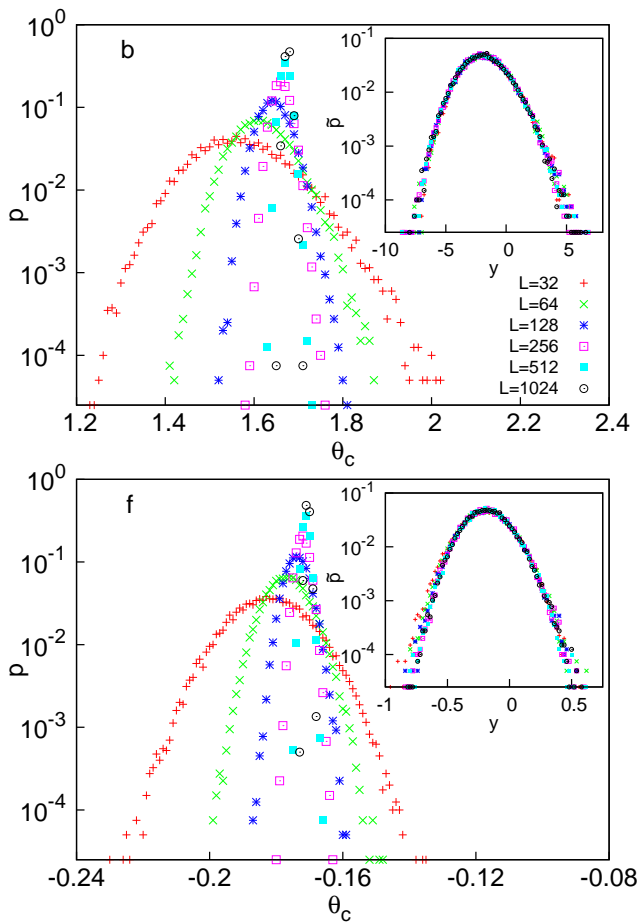}
\end{center}
\caption{
\label{fig_4} (Color online)
Distribution of the pseudo-critical points, $\theta_c(L)$, for various lengths for box-$h$ randomness (upper panel)
and for fixed-$h$ randomness (lower panel). In the insets the scaled distributions are shown
as a function of $y=(\theta_c(L)-\theta_c)L^{1/\nu}$, see the text.
}
\end{figure}

We have calculated pseudo-critical points of square-shaped samples for various linear sizes, $L$, which are expressed
as $L=2^l$ and $L=3*2^{l-1}$ up to $l=10$. Generally we have considered $4\times 10^4$ realizations except for the largest
system, when we had at least $10^4$ samples. The distribution of the pseudo-critical points is shown in Fig.\ref{fig_4}
for both type of disorder. The mean value of the critical points is considerably larger for box-$h$ randomness
and also the width of the distribution - for the same value of $L$ - is larger in this case. (We note that the same trend
is seen in {\it 1d}, where $\theta_c^{(b)}=0$ and $\theta_c^{(f)}=-1$.) Taking into account the result in Eq.(\ref{nu_s})
the appropriate scaling combination is $y=(\theta_c(L)-\theta_c)L^{1/\nu}$ in terms of which the scaled distributions,
$\tilde{p}(y)$,
are shown in the insets of Fig.\ref{fig_4}. Here using our final estimates in Eqs.(\ref{true_cp}) and (\ref{est_nu})
we obtain excellent scaling collapse of the data points for both type of randomness. The scaling curves for the
two different disorder approach the same standardized master curve, which indicates that the fixed point of the
RG transformation is unique and (at least for strong enough disorder) strongly attractive. The master curve is different
from that in {\it 1d}, which is Gaussian in this case\cite{ilrm}. In ${\it 2d}$ the maximum of the curve is shifted to negative values and the
distribution is non-symmetric. We have calculated the percolation (or spanning) probability, $P_{pr}$, at the critical
point, which is given by the fraction of samples having finite replica correlation function at $\theta_c$. It can be
expressed with the scaled distribution function as $P_{pr}=\int_0^{\infty} \tilde{p}(y) {\rm d} y$. Our estimate is:
\be
P_{pr}=0.149(2)\;,
\label{P_pr}
\ee
which is much smaller, than for standard ${\it 2d}$ percolation\cite{perco}.
We have also calculated the skewness, $s$, of the distribution, which has the value $s=0.19(3)$
for both type of disorder. The asymmetric form of the distribution indicates that in ${\it 2d}$ the topology of the renormalized
model is different from that in {\it 1d}. Samples having more strongly connected clusters, thus a higher $\theta_c(L)$,
have a somewhat larger weight than the less strongly connected clusters.

\subsubsection{Shift of the finite-size critical points}
\label{sec:shift}
\begin{figure}[h!]
\begin{center}
\includegraphics[width=3.4in,angle=0]{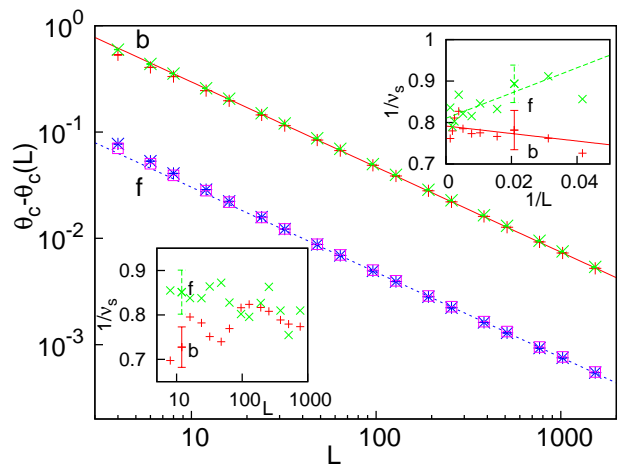}
\end{center}
\caption{
\label{fig_5} (Color online)
Scaling of the shift of the finite-size transition points, calculated from the mean value ({\color{red} $+$} for 'b' and {\color{blue}$\rlap{$\times$}{+}$}
for 'f'), as well as from the median ({\color{green}$\times$} for 'b' and {\color{magenta}$\boxdot$} for 'f') of
the distribution, as a function of $L$ in log-log scale for both type of disorder. Estimates for $\theta_c$ are
taken from Eq.(\ref{true_cp}) and the straight lines indicating the asymptotic behavior have the same
slope: $-1/\nu_s \approx -0.8$. The effective critical exponents calculated from Eq.(\ref{eff_shift}) are
shown in the upper inset (for the mean) and in the lower inset (for the median).}
\end{figure}

For a fixed linear size, $L$, we have calculated the mean value, $\overline{\theta_c}(L)$, as well as the median
$\theta_c^{\rm med}(L)$ of the pseudo-critical points. Since the distribution is non-symmetric these two characteristic
values are not identical, however both are expected to follow the scaling form in Eq.(\ref{nu_s}) with the same
value of the shift exponent, $\nu_s$. This is illustrated in Fig.\ref{fig_5}, in which $\theta_c-\overline{\theta_c}(L)$
as well as $\theta_c-\theta_c^{\rm med}(L)$ is shown as a function of $L$ in a log-log scale for both type of disorder.
Having appropriate limiting values for $\theta_c$, see Eq.(\ref{true_cp}), the points in Fig.\ref{fig_5}
are asymptotically very well on straight lines (both for the mean value and for the median) having approximately the
same slopes: $-1/\nu_s \approx -0.8$.

To get more quantitative estimates we have calculated effective, size dependent shift exponents which are defined as:
\be
\dfrac{1}{\nu_s(L)}=-\dfrac{1}{\ln 2}\ln \left[ \dfrac{\overline{\theta_c}(2L)-\overline{\theta_c}(L)}{\overline{\theta_c}(L)
-\overline{\theta_c}(L/2)}\right]\;,
\label{eff_shift}
\ee
and similarly for the median values. These are shown in the two insets of Fig.\ref{fig_5} as a function of $1/L$
(upper inset for the mean) and $\ln L$ (lower inset for the median), respectively. The exponents calculated
from the mean values show $1/L$ correction terms for both type of disorder, although with different signs.
The extrapolated values are $1/\nu_s^{(b)}=0.79(2)$ and $1/\nu_s^{(f)}=0.81(2)$, which agree within the error
of the calculation and leads the estimate: $1/\nu_s=0.80(2)$. As seen in the lower inset of Fig.\ref{fig_5}
the effective exponents calculated from the median of the distribution have weaker $1/L$ dependence, instead
they show log-periodic-like variations. The estimates for $1/\nu_s$ from these data are compatible with the
previous estimates obtained from the mean values, thus we can write our estimate of the shift exponent:
\be
\nu_s=1.25(3)\;.
\label{est_nu_s}
\ee
%

\subsubsection{Scaling of the width of the pseudo-critical points}
\label{sec:width}
\begin{figure}[h!]
\begin{center}
\includegraphics[width=3.4in,angle=0]{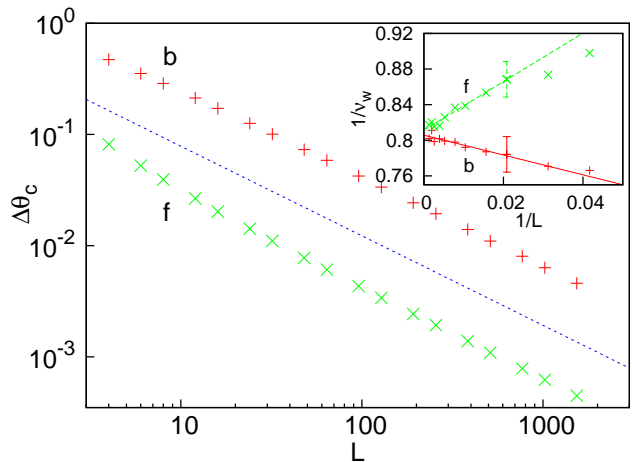}
\end{center}
\caption{
\label{fig_6} (Color online)
Standard deviation of the distribution of the pseudo-critical points as a function of the size in log-log plot
for box-$h$ (upper points) and fixed-$h$ (lower points) randomness. The dotted (blue) straight line has a slope:
$1/\nu_w=-0.808$ corresponding to the estimated value. Inset: finite-size estimates for the exponent, $1/\nu_w$,
plotted as a function of $1/L$.}
\end{figure}

\begin{figure}[h!]
\begin{center}
\includegraphics[width=3.4in,angle=0]{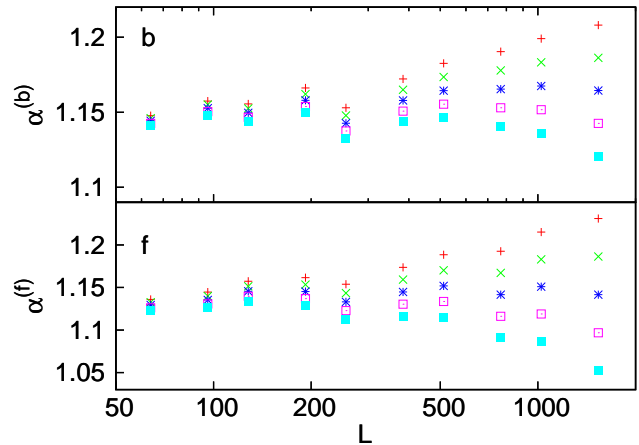}
\end{center}
\caption{
\label{fig_7} (Color online)
The ratio in Eq.(\ref{alpha}) as a function of $\ln L$ for different input values of the critical
point, $\theta_c$. At the 'true' critical point $\alpha(L)$ is approximately $L$ independent.
Upper panel: box-$h$ randomness and $\theta_c$ varies equidistantly between $1.6786$ and $1.6782$ from up to down;
lower panel: fixed-$h$ randomness and $\theta_c$ varies equidistantly between $-0.17030$ and $-0.17038$ from up to down.
}
\end{figure}

We have measured the standard deviation of the distribution of the pseudo-critical points, $\Delta \theta_c(L)$, which are
shown in Fig. \ref{fig_6} as a function of the linear size, $L$, in a log-log scale. In agreement with the scaling
prediction in Eq.(\ref{nu_w}) the points in Fig.\ref{fig_6} are asymptotically on straight lines, the slope of which
is approximately the same for both types of disorder and can be well fitted as $-1/\nu_w \approx -0.8$. As for the
shift exponent in Sec.\ref{sec:shift} we have measured effective, size-dependent critical exponents, which are
defined as:
\be
\dfrac{1}{\nu_w(L)}=\sinh^{-1}\left[- \dfrac{\Delta \theta_c(2L)-\Delta \theta_c(L/2)}{2\Delta \theta_c(L)}\right] \dfrac{1}{\ln 2}\;,
\label{eff_width}
\ee
and plotted in the inset of Fig.\ref{fig_6} as a function of $1/L$. Extrapolating the effective exponents yields $1/\nu_w^{(b)}=0.805(10)$ and $1/\nu_w^{(f)}=0.811(10)$ for the box-$h$ and the fixed-$h$ randomness, respectively.
These values
indeed agree within the error of the calculation, thus we can conclude that $1/\nu_w=0.808(10)$ and thus
\be
\nu_w=1.24(2)\;.
\label{est_nu_w}
\ee
Comparing our estimates for the shift-exponent in Eq.(\ref{est_nu_s}) with that of the width-exponent in Eq.(\ref{est_nu_w})
we notice that they agree within the error of the method, which corresponds to the renormalization group result for
a classical conventional random critical point\cite{aharony}. In order to make a direct check of the equivalence of the two exponents
we have formed the ratio:
\be
\alpha(L)=\dfrac{\theta_c-\overline{\theta_c}(L)}{\Delta \theta_c(L)}\;,
\label{alpha}
\ee
which should approach an $L$-independent constant value at the 'true' critical point, $\theta_c$, provided $\nu_s=\nu_w$.
In Fig.\ref{fig_7} we have plotted the $\alpha(L)$ ratios as a function of $\ln L$ using different input values for
the critical point, $\theta_c$. As one can see in this figure the $L$-dependence of $\alpha(L)$ is very sensitive
to the input value of $\theta_c$, both for box-$h$ and fixed-$h$ randomness, but at its right value the $\alpha(L)$ ratios are approximately $L$ independent. In this way we have demonstrated, that the infinite disorder fixed point of the {\it 2d}
RTFI model is characterized by one correlation-length exponent, which is given by:
\be
\nu=1.24(2)\;.
\label{est_nu}
\ee
Furthermore the ratio in Eq.(\ref{alpha}) at the critical point has the universal value: $\alpha=1.15(2)$, which
does not depend on the form of the randomness. Finally, by this method we have obtained accurate estimates
for the 'true' critical points, which are given by:
\beqn
\theta_c^{(b)}&=&1.6784(1)\; \nonumber \\
\theta_c^{(f)}&=&-0.17034(2)\;.
\label{true_cp}
\eeqn
We have checked, that the values in Eq.(\ref{true_cp}) are consistent with other estimates, which can be obtained
by extrapolating the $\overline{\theta_c}(L)$ data through
Eq.(\ref{nu_s}), but the error bars are smaller. For the box-$h$ disorder the known estimates
are $\theta_c^{(b)}=1.676(5)$, in Ref.[\onlinecite{ladder}] and $\theta_c^{(b)}=1.680(5)$, in Ref.[\onlinecite{yu07}],
which are consistent with that in Eq.(\ref{true_cp}), however the present value has much smaller uncertainty.

\section{Scaling at the critical point}
Having accurate estimates of the critical points for both types of randomness we are ready to study the
critical behavior of the system. In this respect we have concentrated our effort at the critical point,
where we have studied the distribution function of the magnetization, as well as that of the (log-)gaps
and calculated critical exponents by finite-size scaling. At the critical point we have considered finite
systems up to a linear length $L=2048$ and studied $4 \times 10^4$ realizations for each sizes.

These investigations are supplemented with numerical studies outside the critical point, both in
the disordered and in the ordered phases. For the magnetization we have studied the scaling regime, defined
as $\delta L^{1/\nu}=\mathcal{O}(1)$. To obtain information
about the dynamics of the system we have studied the scaling behavior of the excitation energies both in
the ordered and in the disordered Griffiths phases and in this investigation we have not restricted ourselves to
the vicinity of the critical point. In the off-critical region we studied random samples with fixed-$h$ randomness
up to a linear length $L=512$ and for $10^4$ realizations.

\label{sec:crit}

\subsection{Magnetization}
\subsubsection{Spin clusters}
\label{sec:sp_clus}
\begin{figure}[h!]
\begin{center}
\includegraphics[width=3.4in,angle=0]{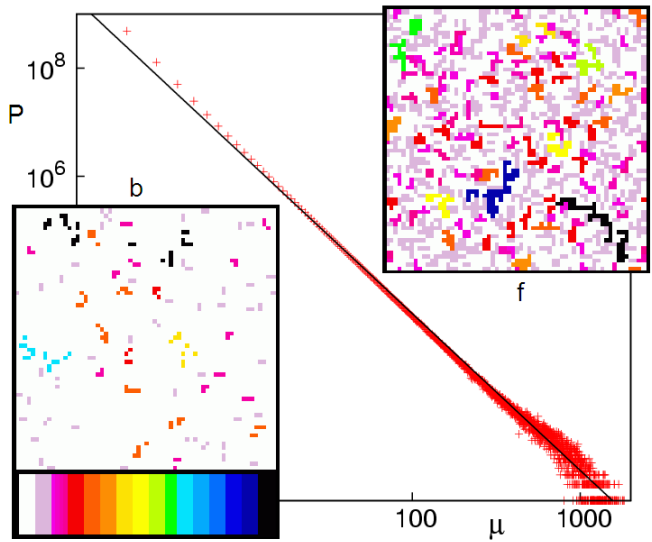}
\end{center}
\caption{
\label{fig_8} (Color online)
Distribution of the mass of the spin clusters in a log-log scale for the $L=1024$ system at the critical
point calculated from 40000 samples. The straight line indicating the asymptotic behavior has a slope $-\tau=-2.9$, see the text.
In the two insets the cluster structure is illustrated
for the $L=64$ system with fixed-$h$ (upper inset) and box-$h$ (lower inset) randomness. The size of the clusters
is increasing with the greyscale (color) as indicated at the bottom of the lower inset.
The one-site clusters are white the points of the largest cluster are black.
}
\end{figure}

During renormalization effective spin clusters are formed, which are than decimated at different energy scales.
We illustrate the cluster structure of a given sample at the critical point in the insets of
Fig.\ref{fig_8} in which clusters having the same size are marked with the same greyscale (color). Note that most of the clusters consist of only one site and the
large clusters are generally geometrically disconnected. We have also analyzed the distribution of the mass of
the clusters, $P_L(\mu)$, which follows the scaling form: $L^{d_f} \tilde{P}(\mu L^{-d_f})$, $d_f$ being
the fractal dimension defined in Eq.(\ref{d_f}). According to scaling theory\cite{perco} the distribution
has a power-law tail, $\tilde{P}(u) \sim u^{-\tau}$, with an exponent $\tau=1+\dfrac{d}{d_f}$. The scaling function
for $L=1024$ is shown in Fig. \ref{fig_8} in a log-log scale and indeed it has a linear dependence with a slope $\tau=2.9(1)$,
which is consistent with our estimate for $d_f$ in Eq.(\ref{d_f_est}).

The magnetic properties of a given finite sample are related to the magnetic moment of some effective
spin cluster, which appears at the last stages of the RG procedure. In principle one can define different types of
such spin clusters. i) The {\it magnetization cluster} has the smallest effective transverse field, thus decimated at the
lowest energy scale. The moment of magnetization clusters are denoted by $\tilde{\mu}$.
 By definition the smallest possible value of the magnetization cluster is $\tilde{\mu}_{min}=1$, thus the corresponding
minimal magnetization at the disordered phase is given by $1/L^d$, i.e. it varies as a power-law of $L$.
ii) The {\it correlation cluster} with a moment $\tilde{\mu}_{corr}$ is present in the duplicated sample and
involved in the replica
correlation function. By definition a correlation cluster exists only for $\theta \le \theta_c(L)$, i.e. below
the pseudo-critical point of the given sample. We have checked, that for large $L$ if a correlation cluster exists,
it is almost always the magnetization cluster.
iii) Finally we define also an {\it energy cluster} in a sample, which is the magnetization cluster for $\theta > \theta_c(L)$,
whereas for $\theta \le \theta_c(L)$ it is the cluster decimated before the correlation cluster. These clusters are involved
in the low-temperature or the small longitudinal field properties of the system and will be considered in dynamical scaling
in Sec.\ref{sec:dyn}.
\subsubsection{Moment of magnetization clusters}
\begin{figure}[h!]
\begin{center}
\includegraphics[width=3.4in,angle=0]{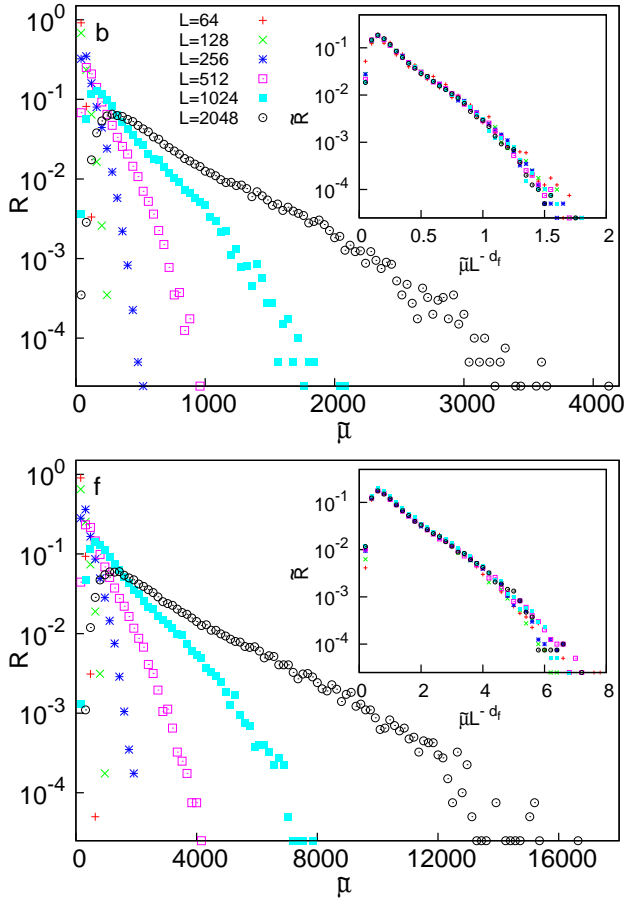}
\end{center}
\caption{
\label{fig_9} (Color online)
Distribution of the moment of magnetization clusters for different lengths. Upper panel: box-$h$ randomness,
lower panel: fixed-$h$ randomness. The scaled distributions are shown in the insets, where for the fractal
dimension the estimate in Eq.(\ref{d_f_est}) is used.
}
\end{figure}

We have calculated the distribution functions of the moments of magnetization clusters for different lengths,
$R_L(\tilde{\mu})$, which are
shown in Fig.\ref{fig_9} for both types of randomness. According to scaling theory
$R_L(\tilde{\mu})=L^{d_f}\tilde{R}(\tilde{\mu} L^{-d_f})$
which is illustrated in the insets of Fig.\ref{fig_9}. Up to a multiplicative constant the scaling functions, $\tilde{R}(\omega)$,
are identical for the two different randomness and can be approximated with an exponential function: $\tilde{R}(\omega) \sim
\exp(-\omega/\omega^*)$, $\omega^*$ being some randomness dependent value.

\subsubsection{Fractal dimension and critical exponent}
\begin{figure}[h!]
\begin{center}
\includegraphics[width=3.4in,angle=0]{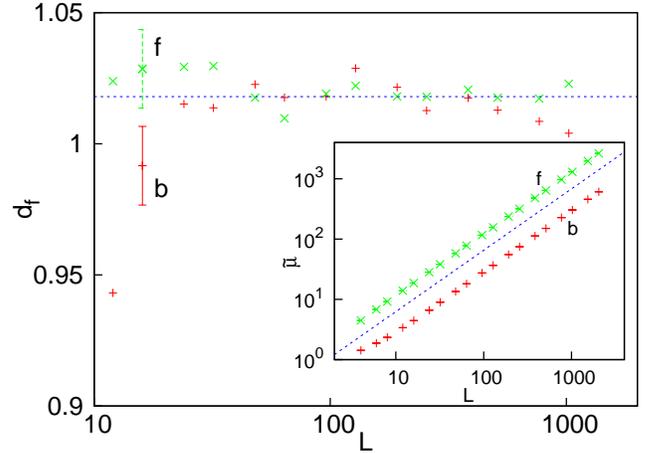}
\end{center}
\caption{
\label{fig_10} (Color online)
Finite-size estimates for the fractal dimension of the magnetization cluster in Eq.(\ref{d_f_eff})
as a function of $\ln(L)$ for the fixed-$h$ (f) and the box-$h$ (b) randomness. The dotted horizontal line
at $d_f=1.018$ represents the mean value and our estimate. In the inset the average moment of the
magnetization clusters are shown as a function of $L$ in a log-log scale for both type of randomness.
The dotted straight line has the slope $d_f$ as extracted from the main figure.}
\end{figure}

In order to obtain an accurate estimate for the fractal dimension, $d_f$, and for the magnetization exponent, $x$,
we have calculated average moments of the magnetization clusters, $\overline{\mu}_L$, which are plotted
in the inset of Fig. \ref{fig_10} as a function of $L$ in a log-log scale. For both type of randomness the points are
asymptotically on straight lines having the same slope, which is in agreement with the scaling relation in
Eq.(\ref{mu}). We have calculated effective, size-dependent fractal dimensions through:
\be
d_f(L)=\sinh^{-1}\left[\dfrac{\overline{\mu}_{2L}-\overline{\mu}_{L/2}}{2\overline{\mu}_{L}}\right] \dfrac{1}{\ln 2}\;,
\label{d_f_eff}
\ee
which are plotted in Fig. \ref{fig_10}. As seen in this figure the effective fractal dimensions show no systematic
trend with $L$ and the $d_f(L)$ values spread around the same mean value for both form of disorder. This mean value
is taken to our estimate for the fractal dimension:
\be
d_f=1.018(15)
\label{d_f_est}
\ee
and from Eq.(\ref{d_f}) we have for the magnetization exponent:
\be
x=0.982(15)\;.
\label{x_f}
\ee
%

\subsubsection{Scaling of the magnetization}
\label{magn_g}
\begin{figure}[h!]
\begin{center}
\includegraphics[width=3.4in,angle=0]{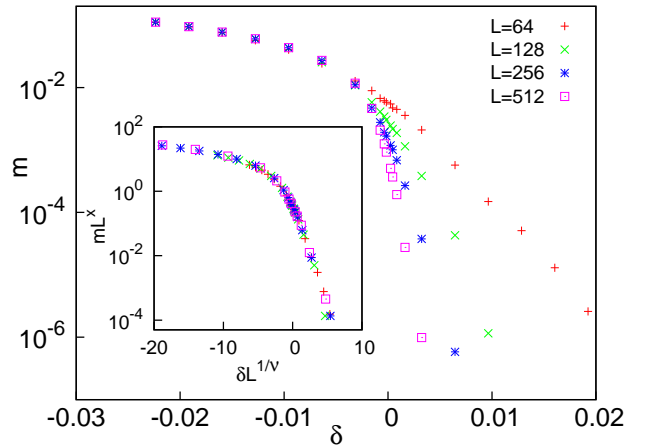}
\end{center}
\caption{
\label{fig_11} (Color online)
Magnetization as a function of the control-parameter in the vicinity of the critical point for different
finite systems. In the inset the scaled magnetization $m(\delta,L)L^{x}$ is plotted
as a function of $\delta L^{1/\nu}$.}
\end{figure}

The magnetization is given by the asymptotic value of the correlation
function in Eq.(\ref{m_def}), which definition can be extended in a finite system
in terms of the replica correlation function as introduced in Sec.\ref{sec:pseudo-cr-pt}.
In a given sample of linear size $L$ two spins at a distance
$r \sim L$ are correlated if both are in the same correlation cluster. Consequently (the average value of)
the magnetization is given by: $m(\delta,L)=\dfrac{\overline{\mu}_{corr}}{L^d}$, where $\overline{\mu}_{corr}$
is the average value of the mass of the correlation cluster as defined in Sec.\ref{sec:sp_clus}. The magnetization as a function
of $\delta$ is plotted in Fig.\ref{fig_11} for different finite systems. Using scaling theory the magnetization
in the vicinity of the critical point is expected to behave as $m(\delta,L)=L^{-x} \tilde{m}(\delta L^{1/\nu})$.
To test this assumption in the inset of Fig.\ref{fig_11} we have plotted $m(\delta,L)L^{x}$ as a function of
$\delta L^{1/\nu}$. Using the estimates for the critical exponents in Eqs.(\ref{est_nu}) and (\ref{x_f}) an excellent
scaling collapse of the data is obtained.

\subsection{Dynamical scaling}
\label{sec:dyn}
Here we study the properties of the low-energy excitations, which are responsible for the
dynamical behavior of the system, such as the auto-correlation function or the low-temperature and
small-field behavior of the susceptibility, specific heat, magnetization, etc.
Such a low-energy collective excitation of a random sample of linear length $L$ in the SDRG method is represented by
a large spin cluster which is formed at the last steps of the renormalization process and its excitation energy, $\epsilon_L$,
is measured by the effective transverse field of the cluster. These clusters are the so called energy clusters as defined
in Sec.\ref{sec:sp_clus}.
According to the scaling relation in Eq.(\ref{psi}) at the
critical point it is convenient to use the log-variable: $\gamma_L=-\ln(\epsilon_L)$.
In the following we study the distribution of $\gamma_L$ at the critical point, as well as in disordered and ordered
Griffiths phases, and investigate its scaling behavior with $L$.  As far as dynamical properties
are considered the SDRG method gives asymptotically exact results also in the off-critical region, where
the relaxation time is divergent\cite{igloi02}. This is not the case for static quantities due to the finite correlation length.

\subsubsection{Critical point}
\begin{figure}[h!]
\begin{center}
\includegraphics[width=3.4in,angle=0]{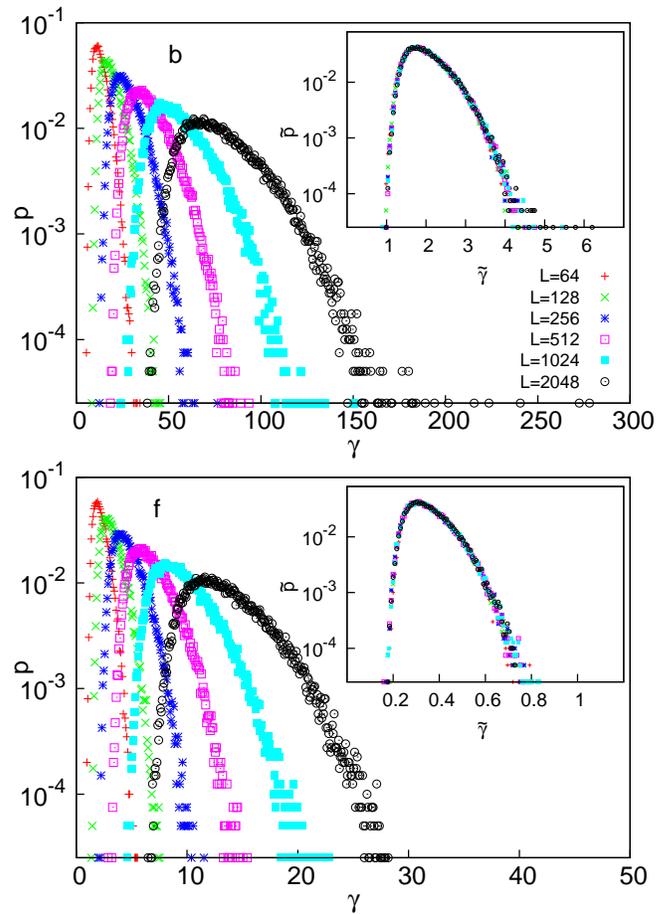}
\end{center}
\caption{
\label{fig_12} (Color online)
Distribution of the log-excitation energies at the critical point for different finite systems. Upper panel:
box-$h$ randomness; lower panel: fixed-$h$ randomness. In the insets scaling collapse of the distributions in terms
of $\tilde{\gamma}=(\gamma-\gamma_0) L^{-\psi}$ are shown, with $\psi = 0.48$. The constant is $\gamma_0=-1.5(4)$
for box-$h$ randomness and $\gamma_0=-0.4(1)$ for fixed-$h$ randomness, respectively.}
\end{figure}

At the critical point the energy clusters are decimated either at the last step of the renormalization process,
which happens
if the pseudo-critical point of the sample, $\theta_c(L)$, is smaller than $\theta_c$, or at the last but one step,
if $\theta_c(L)>\theta_c$.
The distribution of the log-excitation energies at the critical point for different sizes are shown in Fig.\ref{fig_12}
for the two type of randomness. As seen in this figure the distributions broaden with increasing $L$, which is a
clear signal of infinite-disorder scaling. Referring to Eq.(\ref{psi}) we introduce the scaling combination:
$\tilde{\gamma}=(\gamma_L-\gamma_0)L^{-\psi}$, in terms of which the distributions collapse to the same curve provided
the exponent is $\psi \approx 0.48$. This is illustrated in the insets of Fig.\ref{fig_12}. The constant term, $\gamma_0$, used
in the fitting process is found to be $\mathcal{O}(1)$ and has only a little influence on the value of the exponent $\psi$.
The scaling functions in the insets of Fig.\ref{fig_12} have the same form for the two types of applied randomness
and have a heavier tail than in 1$d$. The skewness values at $L=1024$ are $s=0.82(1)$ for {\it 2d} to be compared with
$s=0.64(1)$ in {\it 1d}.

\begin{figure}[h!]
\begin{center}
\includegraphics[width=3.4in,angle=0]{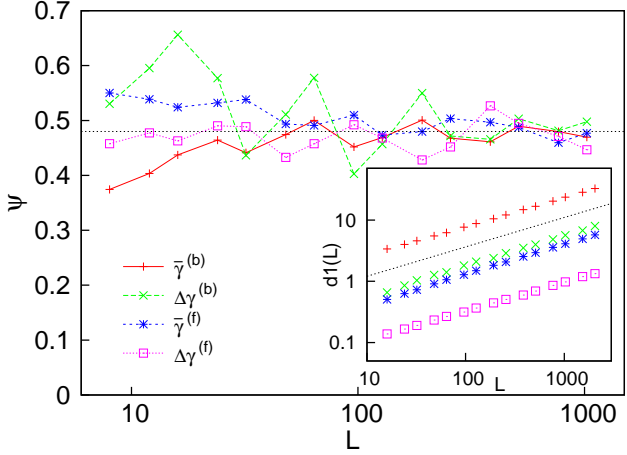}
\end{center}
\caption{
\label{fig_13} (Color online)
Effective exponents $\psi(L)$ as a function $\ln(L)$ calculated from the mean or from the width of the distribution
of the log-excitation energies for the two type of randomness. Inset: finite differences of the mean and the width of the log-excitation energy distribution as a function of $L$ in a log-log scale, see text. The slope of the dotted straight line
is given by $\psi=0.48$.}
\end{figure}

In order to obtain more accurate estimate for the exponent $\psi$ we have considered the mean value, $\overline{\gamma}(L)$,
as well as the standard deviation, $\Delta \gamma(L)$, of the distributions. We have formed the differences, $d1(L)$,
which are $d1(L)=\overline{\gamma}(L)-\overline{\gamma}(L/2)$ for the mean and
$d1(L)=\Delta \gamma(L)-\Delta \gamma(L/2)$ for the width of the distribution, respectively and which are plotted
in the inset of Fig.\ref{fig_13} as a function of $L$ in a log-log plot. As seen in this figure the points for the
two quantities and for the two-type of randomness are on parallel straight lines the slope of which is compatible
with $\psi \approx 0.48$. In the next step we have calculated finite-size effective exponents through the definition:
\be
\psi(L)=\dfrac{1}{\ln 2}\ln \left[ \dfrac{d1(2L)}{d1(L)}\right]\;,
\label{eff_psi}
\ee
which are plotted in Fig.\ref{fig_13} as a function of $L$. As seen in this figure there is
some systematic trend of the points up to $L=128-192$, but for larger $L$-s there are only fluctuations around a mean value
what we use as an estimate for the exponent:
\be
\psi=0.48(2)\;.
\label{est_psi}
\ee
This limiting value is the same for both type of randomness and describes well the scaling behavior both the
mean value and the width of the distribution.

\subsubsection{Disordered Griffiths phase}
\begin{figure}[h!]
\begin{center}
\includegraphics[width=3.4in,angle=0]{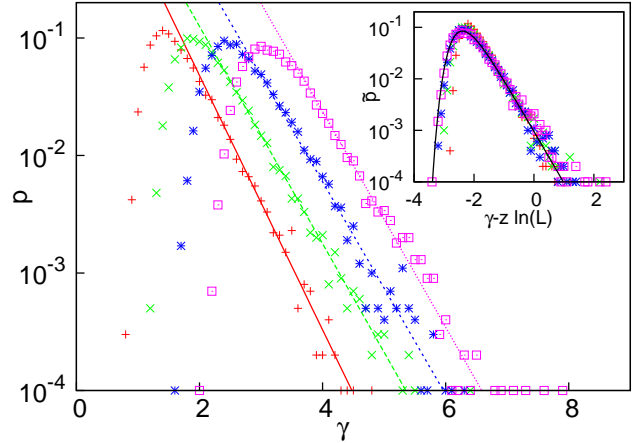}
\end{center}
\caption{
\label{fig_14} (Color online)
Distribution of the log-excitation energies in the disordered Griffiths-phase (at $\delta=9.64\times 10^{-3}$)
in a log-lin scale for different sizes. The slopes of the straight lines indicating the tail of the curves are
$d/z=2.5(2),~2.1(2),~2.1(2)$ and $2.1(2)$, for $L=64,128,256$ and $512$, respectively. In the inset the scaled distributions
are shown in terms of the variable in Eq.(\ref{gamma_0}) with $d/z=2.3(2)$, which is well described by the Fr\'echet distribution
(full line) in Eq.(\ref{extr_value}).}
\end{figure}

In the disordered Griffiths phase the energy clusters, which are related to the low-energy excitations
in a large system are decimated at the last step of the renormalization process.
The scaling behavior of the low-energy excitations here is different from that at the critical point.
We illustrate the scaling behavior of the distribution of the log-excitation energies for different
sizes in Fig.\ref{fig_14} for the fixed-$h$ randomness at a distance $\delta=9.64 \times 10^{-3}$ from
the critical point. As seen in this figure the shape of the (logarithm of the) distribution functions is very similar
for different $L$-s and the curves are merely shifted with $\ln(L)$. This follows from the assumption, that the typical
value of the excitation energy in a system of size $L$ scales as a power-law:
\be
\epsilon_L \sim L^{-z}\;,
\label{exp_z}
\ee
$z$ being the dynamical exponent, which is a continuous function of the control parameter, $\delta>0$\cite{disord_scaling}.
Consequently the appropriate scaling combination is:
\be
\tilde{\gamma}=\gamma_L-z \ln(L)-\gamma_0\;,
\label{gamma_0}
\ee
in terms of which the distribution functions have a scaling collapse, provided the appropriate value
of the dynamical exponent is used. This is illustrated in the inset of Fig.\ref{fig_14}. An estimate for
the dynamical exponent from the shift of the distributions, $z_{sh}(L)$, can be obtained from the
optimal collapse of the data points for sizes $L/2$ and $L$. In Fig.\ref{fig_14} we have $d/z_{sh}(512)=2.3(2)$.
If the low-energy excitations in the system are localized then 
the scaled distribution function is suggested\cite{jli06} to be given by the Fr\'echet distribution known from extreme value statistics\cite{galambos}
in the form:
\be
\ln p(\tilde{\gamma}-\gamma_0)=-\dfrac{d}{z}\tilde{\gamma}-\exp\left(-\tilde{\gamma}\dfrac{d}{z}\right)+\ln(d/z)\;.
\label{extr_value}
\ee
Indeed the scaled distribution function in the inset of Fig.\ref{fig_14} is well described by the function
in Eq.(\ref{extr_value}), where only one fitting parameter, $\gamma_0$ in Eq.(\ref{gamma_0}) is used.
We note that for large $\tilde{\gamma}$ the tail of $\ln p(\tilde{\gamma})$ is linear and its slope, $-d/z_{sl}(L)$,
can be used to obtain an independent estimate for the dynamical exponent. The measured values of $d/z_{sl}(L)$
are given in the caption of Fig.\ref{fig_14} and these are compatible with those calculated from the shift of the
distributions. There are, however finite-size corrections for $L<\xi$, where the correlation length
at the studied value of $\delta$ is of the order of $\xi=\mathcal{O}(10^2)$.

\begin{figure}[h!]
\begin{center}
\includegraphics[width=3.4in,angle=0]{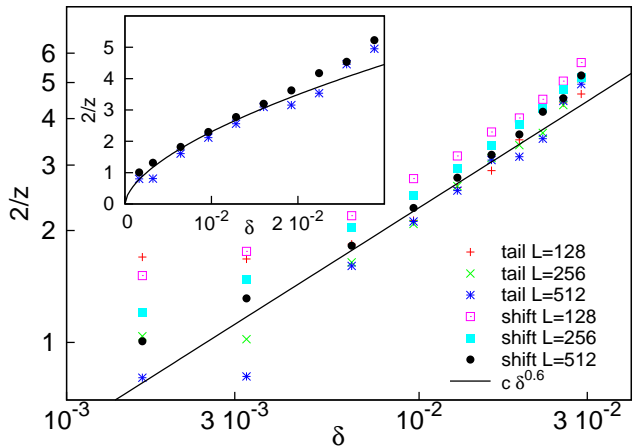}
\end{center}
\caption{
\label{fig_15} (Color online)
Estimates for $2/z$ at different points of the disordered Griffiths phase in a log-log plot. The estimates are calculated
either from the shift of the distributions or from their tail at various finite sizes. The straight line with a slope $0.6$ indicates the asymptotic behavior as given in Eq. (\ref{zscal}). In the inset data for the largest system, $L=512$, are
compared with the scaling curve in Eq. (\ref{zscal}) with $c=36.5$ in linear plot.}
\end{figure}

We have repeated the previous calculation for several values of $\delta$ in the disordered Griffiths phase and
calculated estimates for the dynamical exponent both from the shift of the distributions and from the slope
of the tail. These estimates obtained at different $L$-s are shown in a log-log plot in Fig. \ref{fig_15}. One
can notice that the finite-size corrections are stronger for small $\delta$-s, where the correlation length
is comparatively larger. According to scaling theory\cite{im} the dynamical exponent for
small $\delta$ behaves as 
\be
\dfrac{d}{z}\approx c \delta^{\nu\psi}\;
\label{zscal}
\ee
and divergent at $\delta=0$. We have checked the relation in Eq.(\ref{zscal}) and indeed in Fig.\ref{fig_15}
one can identify an approximately linear part for $\delta \le 0.02$ having a slope $\approx 0.6$. This value
is compatible with our previous estimates $\nu\psi=0.60(6)$ using results in Eqs.(\ref{est_nu}) and (\ref{est_psi}).

Closing this section we note that the dynamical exponent enters into the singularities of different physical
quantities. For example at low-temperature the susceptibility behaves as: $\chi(T) \sim T^{d/z-1}$ and the specific
heat has the form $C_v \sim T^{d/z}$. More details about the scaling relations in the Griffiths phase can
be found in\cite{im,vojta06}.

\subsubsection{Ordered Griffiths-phase}
\begin{figure}[h!]
\begin{center}
\includegraphics[width=3.4in,angle=0]{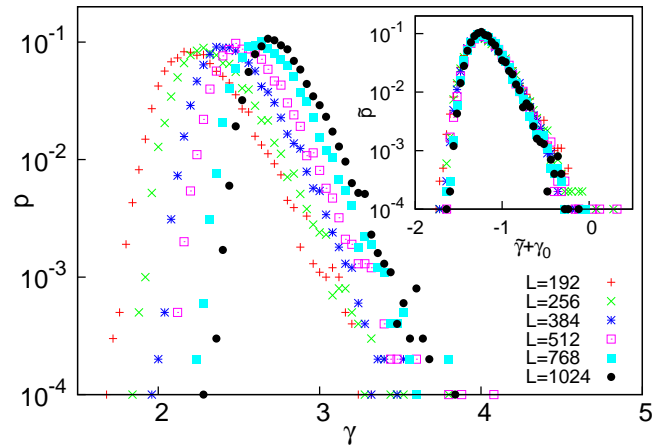}
\end{center}
\caption{
\label{fig_16} (Color online)
Distribution of the log-excitation energies in the ordered Griffiths-phase (at $\delta=-1.916\times 10^{-2}$)
in a log-lin scale for different sizes. In the inset the scaled distributions
are shown in terms of the variable in Eq.(\ref{gamma_ord}).}
\end{figure}

In the ordered phase of a large system with $\delta<0$ there is an infinite magnetization cluster, which is
decimated at the last step of the renormalization procedure. The low-energy excitations of the system here
are related to the energy cluster, which has a finite extent and which is decimated just before the magnetization
cluster. The distribution of the log-excitation energies for different sizes are shown in Fig.\ref{fig_16}
for the fixed-$h$ randomness at $\delta=-1.916\times 10^{-2}$. Comparing the distributions with that of
the disordered Griffiths phase in Fig.\ref{fig_14} one can notice that in both cases the distributions
are not broaden but shifted with an $L$-dependent amount. There are, however, several differences in the two
figures. In the ordered Griffiths phase the finite-size effects are stronger, therefore we went up to $L=1024$.
More importantly, the shift of the distributions in the ordered Griffiths phase is slower than linear with $\ln(L)$.
This is connected to the scaling result, that the typical value of the excitation energy, $\epsilon_L$, is related to
the size of the system as: $\ln \epsilon_L \sim -\ln^{1/d}(L)$, thus the appropriate scaling combination is\cite{ord_scaling}
\be
\tilde{\gamma}=\gamma_L-A \ln^{1/d}(L)-\gamma_0\;,
\label{gamma_ord}
\ee
which is to be compared with Eq.(\ref{gamma_0}). Indeed using the variable in Eq.(\ref{gamma_ord}) the distributions
show a scaling collapse, as illustrated in the inset of Fig.\ref{fig_16}.

Also the shape of the scaled distributions are different in the two Griffiths phases. In the disordered
Griffiths phase the distributions in the inset of Fig.\ref{fig_14} approach a linear asymptotics from above,
on the contrary in the ordered Griffiths phase in the inset of Fig.\ref{fig_16} the points bend below
a straight line. This is compatible with the scaling result, that asymptotically\cite{ord_scaling}: 
\be
\ln p(\tilde{\gamma}) \sim -\tilde{\gamma}^d\;.
\label{extr_value1}
\ee
Our data in the inset of Fig.\ref{fig_16} are still not in the asymptotic regime but the tail of the
distribution clearly decreases faster than linear for the large sizes.

\section{Discussion}
\label{sec:disc}
The concept of infinite disorder fixed point has been introduced quite recently\cite{danielreview} and its basic
properties have been demonstrated in partially exact calculations in different one-dimensional
random quantum\cite{mccoywu,
shankar,fisher,senthil,carlon-clock-at} and stochastic systems\cite{rgsinailong,hiv,igloipartial}.
In higher dimensions, in particular in two dimensions the
calculations are numerical and have only limited accuracy\cite{motrunich00,lin00,karevski01,lin07,yu07,ladder,pich98}. In the present paper we have considered
the prototypical model in {\it 2d} having an IDFP the random transverse-field Ising model and studied its
critical properties by a numerical implementation of the SDRG approach. As follows from the concept
of IDFP our numerical results are expected to be asymptotically exact.

In our approach we have used a very efficient computational algorithm which made us possible to treat samples which
are ten-times larger in linear size, compared with previous calculations. In this way we could reduce the
effect of finite-size corrections and could also study off-critical properties, such as scaling functions
and dynamical scaling in the disordered and ordered Griffiths phases.

The main results of our investigations are the following. We have extended the finite-size scaling study
for pseudo-critical points and from their distribution we have obtained precise estimate for the
correlation length critical exponent, which has been shown to govern both the shift and the width of
the distribution. Using different types of randomness we have demonstrated that the IDFP is universal,
the critical exponents as well as the critical scaling functions are independent of the randomness
used in the calculation. We have also studied cross-over phenomena with respect of the linear size of
the system as well as the type of randomness used in the calculation. The sometimes large errors and
deviations between the results of previous numerical studies are presumably due to cross-over effects.
This can be seen, e.g. in Fig.\ref{fig_13} where the estimates of the exponent $\psi$ have strong finite-size
as well as randomness-type dependence.

As a result of the larger samples and the good statistics of the numerical data we have obtained
accurate estimates for the critical exponents and studied - at the first time - the behavior of
scaling functions, both at the critical point and in the finite-size scaling limit in the vicinity
of the critical point. We have also extended our investigations to the disordered and the ordered
Griffiths phases and have checked various predictions of phenomenological scaling theory.

Comparing the critical behavior in {\it 1d} to that in {\it 2d} we have qualitatively similar results, but
there are also important differences. First of all
the actual values of the critical exponents as well as the form of the scaling functions are different.
Scaling in the ordered Griffiths-phase, which involves powers of $d$ in Eqs.(\ref{gamma_ord}) and
(\ref{extr_value1}) however, is also qualitatively different. One particular feature of the model in
{\it 1d} is self-duality, which could be the reason why 
in {\it 1d} the distribution of the pseudo-critical points involves different shift and mean exponents\cite{ilrm}.

Our calculations can be extended to several directions. One possibility is to study the entanglement
properties of the model\cite{Amicoetal08} which can be well performed within the frame of the SDRG approach\cite{refael}.
At present
conflicting theoretical predictions are available about the finite-size dependence of the critical
entanglement entropy\cite{lin07,yu07}, which could be possibly clarified by using larger systems in the calculation.
A second promising direction of application of our approach is to consider higher dimensional
systems. At present even in three dimensions only the existence of infinite disorder scaling is demonstrated\cite{motrunich00}, but
no estimates are known about the critical exponents. In four and higher dimensions no numerical
studies have been performed so far. Studies in these directions are in progress.

\begin{acknowledgments}
This work has been supported by the Hungarian National Research Fund under grant No OTKA
K62588, K75324 and K77629 and by a German-Hungarian exchange program (DFG-MTA).
We thank to N. Kawashima for communicating us a preliminary version of the proof of the filtering condition.
We are grateful to P. Sz\'epfalusy and H. Rieger for useful discussions.
\end{acknowledgments}

\appendix
\section{Filtering out irrelevant bonds}
In the SDRG procedure if the maximum rule is applied several bonds are irrelevant, which means that they
can be deleted without modifying the final result of the renormalization.
\label{kawa_app}
\subsection{Condition for a bond to be irrelevant}
Let us consider a bond with a log-coupling $\kappa_{ij}=\ln J_{ij}$ between sites $i$ and $j$ and consider such
nearest neighbor points, one of those is denoted by $k$ in Fig. \ref{fig_17}, which has bonds both to $i$ and $j$ of strength
$\kappa_{ik}$ and $\kappa_{jk}$, respectively. The triangle $(i,j,k)$ is called the
'majorating triangle' of the  $(i,j)$ bond if $\kappa_{ij}$ is the smallest bond in the triangle,
and it is also smaller, than the potentially generated new bond: $\kappa'_{ij}=\kappa_{ik}+\kappa_{jk}-\theta_k$,
where $\theta_k=\ln h_k$ is the log-transverse field at site $k$.
If such a majorating triangle exists then the bond is irrelevant.

\subsection{Proof of the filtering criterion}
\label{proof}
\begin{figure}[h!]
\begin{center}
\includegraphics[width=1.5in,angle=0]{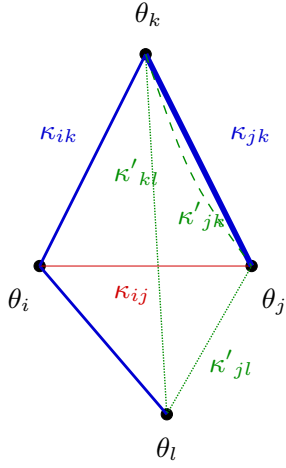}
\end{center}
\caption{
\label{fig_17} 
Illustration of the neighborhood of an irrelevant bond $(i,j)$ having a majorating triangle $(i,j,k)$.
The values of the log-couplings and the log-transverse fields are also indicated, see the text.
}
\end{figure}

For the proof it is enough to consider such a decimation step during which the majorating triangle collapses and
we follow also the evolution of the parameters in
another triangle, $(i,j,l)$, which is not necessarily a majorating one. The maximal log-parameter in the
four-site system in Fig. \ref{fig_17} is denoted by $\omega$. We note that during renormalization the log-transverse
fields can not increase, thus the change of their values are unimportant in the proof.

\subsubsection{Bond decimation}
The $(i,j)$ bond is not the largest one, since it has a majorating triangle, so that for the position of
the largest bond we have three different cases.
i) Largest bond in the
majorating triangle: $\omega=\kappa_{ik}$ (or $\kappa_{jk}$). The spins $i$ and $k$ fuse into a new effective spin, which is connected to the spin $j$ with a coupling of  $\kappa'=\max{(\kappa_{ij}, \kappa_{jk})}=\kappa_{jk}$, thus, $\kappa_{ij}$ simply disappears. ii) Largest bond in the another triangle: $\omega=\kappa_{jl}$ (or $\kappa_{il}$). The $j$ and $l$ spins fuse into a new effective spin. If $\kappa_{ij}<\kappa_{il}$, than the value of $\kappa_{ij}$ simply disappears, otherwise its value does not change. Similarly, the value of $\kappa_{ik}$ does not change, while $\kappa_{jk}$ is replaced by $\kappa_{jk}'=\max{(\kappa_{jk},\kappa_{kl})}\geq\kappa_{jk}$. Taken all round, the new $(i,j,k)$ triangle is a majorating triangle of the $(i,j)$ bond. iii) Largest bond between the two triangles: $\omega=\kappa_{kl}$. The spins
$k$ and $l$ fuse and there is one triangle left. Here the value of $\kappa_{ij}$ does not change and the new
triangle is a majorating one of the $(i,j)$ bond.

\subsubsection{Site decimation}
We should consider two different cases: i) $\omega=\theta_{k}$: $\kappa_{ij}'=\kappa_{ik}+\kappa_{jk}-\theta_{k}>\kappa_{ij}$ due to our assumptions, thus $\kappa_{ij}$ disappears without affecting the results. ii) $\omega=\theta_{i}$ (or $\theta_{j}$ analogously): $\kappa_{jk}'=\kappa_{ij}+\kappa_{ik}-\theta_{i}\leq\kappa_{ij}<\kappa_{jk}$, thus this coupling is unaffected by the $\kappa_{ij}$ coupling. However, in principle the other newly generated bonds are not irrelevant, e.g. $\kappa_{jl}'<\kappa_{jl}$ is not always fulfilled. Due to this, the effect of the majorated $(i,j)$ edge will not certainly disappear after one single decimation step, in which it is involved. Moreover, a lot of new couplings can be generated
occasionally, if the bond lies in a 'dangerous position'.

\subsubsection{Completing the proof}
Here we show, that the newly generated coupling in the $(l,j,k)$ triangle, $\kappa_{jl}'$, is always majorated by this
triangle. The two neighboring couplings of $\kappa_{jl}'$ are $\kappa_{jk}$ (due to the fact, that $\kappa_{jk}'<\kappa_{jk}$), and $\max{(\kappa_{kl},\kappa_{kl}')}$, where $\kappa_{kl}'=\kappa_{ik}+\kappa_{il}-\theta_{i}$. $\kappa_{jl}'$ is smaller than these neighboring couplings, namely $\kappa_{jl}'<\kappa_{ij}<\kappa_{jk}$ and $\kappa_{kl}'-\kappa_{jl}'=\kappa_{ik}-\kappa_{ij}>0$. From the latter follows, that $\kappa_{jl}'<\kappa_{kl}'$, which can not be greater, than $\max{(\kappa_{kl},\kappa_{kl}')}$ corresponding to the new value of the coupling between the $k$ and $l$ spins. Now we see, that the $\kappa_{jl}'$ coupling is majorated by the $(l,j,k)$ triangle, if $\kappa_{jl}'$ is smaller, than the generated
coupling $\kappa_{jl}''\equiv\max{\left(\kappa_{kl},\kappa_{kl}'\right)}+\kappa_{jk}-\theta_k$, which is obtained by
decimating the transverse field at $k$. Let us consider their difference: $\kappa_{jl}''-\kappa_{jl}'\geq \kappa_{kl}'+\kappa_{jk}-\theta_k-\kappa_{ij}-\kappa_{il}+\theta_{i}=\kappa_{ik}+\kappa_{jk}-\theta_k-\kappa_{ij}=\kappa_{ij}'-\kappa_{ij}>0$. Thus the $\kappa_{jl}'$ coupling is always majorated by the $(l,j,k)$ triangle.

Let us summarize our findings. If the bond $(i,j)$ has a majorating triangle, it can not be decimated directly.
Decimating in its neighborhood either this bond disappears or new couplings are generated, the value of which involves
$\kappa_{ij}$. These new couplings, however, are always majorated by a triangle, thus they are never decimated during the
renormalization process. Consequently the value of $\kappa_{ij}$ does not influence the result of the renormalization.


\begin{thebibliography}{99}
\bibitem{note} There is a third advantage of the maximum rule, which follows from the fact that in this case the average
couplings are weaker than for the summing rule. As a consequence the critical point of the system is considerably
lower than for the summing rule and therefore less $h$-decimations have to be performed in the vicinity of the critical
point, which speeds up the calculation.
\bibitem{disord_scaling}
The relation in Eq.(\ref{exp_z}) is usually explained\cite{im} that a low-energy excitation is associated to
a rare region of size $\ell$, having $\sim \ell^d$ strong bonds and being locally in the ordered phase.
Such a rare region appears with an exponentially small probability: $p(\ell) \sim \exp(-\alpha \ell^d)$ and the
corresponding excitation energy is also exponentially small: $\epsilon(\ell) \sim \exp(-\sigma \ell^d)$. The
typical size of the largest rare region, $\ell^*$, in a finite system with $L^d$ sites follows from the relation:
$L^d p(\ell^*)=\mathcal{O}(1)$, thus the smallest excitations are typically:
$\epsilon_L=\epsilon(\ell^*) \sim L^{-z}$, with $z=\dfrac{\sigma}{\alpha}d$.
\bibitem{ord_scaling}
The reasoning in [\onlinecite{disord_scaling}] should be modified, by taking account that the ordered rare region of
size $\ell$ should be separated from the infinite cluster with a region of weak bounds having a distance of $\ell_0 \sim \ell^d$.
(Otherwise the rare region will be a renormalized as a part of the infinite cluster.) The existence of a separated
rare region has the probability, $p(\ell) \sim \exp(-\alpha' \ell_0^d) \sim \exp(-\alpha \ell_0^{d^2})$, whereas
its excitation energy is $\epsilon(\ell) \sim \exp(-\sigma \ell^d)$. Following the reasoning in\cite{disord_scaling}
we arrive to the result in Eq.(\ref{gamma_ord}). The form of the tail of the distribution in Eq.(\ref{extr_value1})
follows similarly.
\end{thebibliography}
\end{document}